\documentclass[aps,showpacs,twocolumn,showkeys,amsmath,amssymb,floatfix,nofootinbib,preprintnumbers]
{revtex4}
\usepackage{bm}

\begin{document}
 
\title{Heavy-Quark Hybrid Mass Splittings: Hyperfine and ``Ultrafine''}

\author{Richard F. Lebed}
\email{richard.lebed@asu.edu}
\affiliation{Department of Physics, Arizona State University, Tempe,
Arizona 85287-1504, USA}

\author{Eric S. Swanson}
\email{swansone@pitt.edu}
\affiliation{Department of Physics and Astronomy, University of
Pittsburgh, Pennsylvania 15260, USA}

\date{August, 2017}

\begin{abstract}
It is argued that the heavy-quark limit of QCD requires a certain
combination of hyperfine mass splittings in heavy-quark hybrid-meson
multiplets to be unusually small.  This observation will assist in the
exploration of the heavy-quark hybrid spectrum at facilities such as
PANDA.  Alternatively, a large measured value for this mass splitting
indicates that at least one member of the multiplet must contain
significant light-quark degrees of freedom.
\end{abstract}

\pacs{14.40.Rt, 12.39.Hg, 12.39.Jh, 14.40.Pq}

\keywords{Exotic hadrons, hybrids, quarkonium}
\maketitle


\section{Introduction} \label{sec:Intro}

The term {\it hyperfine splitting\/} refers to the difference between
two energy levels of a particle, atom, or molecule due to interactions
involving a coupling to the magnetic dipole or higher electromagnetic
moments of a nucleus.  In atoms or molecules, hyperfine splittings are
suppressed due to the large size of the nuclear mass $m_N$ compared to
that of the electrons, either because $1/m_N$ appears in the nuclear
magnetic dipole moment, or due to large gradients of internal atomic
electromagnetic fields in the vicinity of the nucleus (within a
Compton wavelength $1/m_N$) that can produce an observable coupling
to the nucleus.

In the case of heavy-quark hadrons, the heavy quark $Q$ assumes the
role of the nucleus, and being spin-1/2, its only higher
electromagnetic moment is the magnetic dipole $\bm{\mu}_Q$, which is
proportional to its spin $\bm{S}_Q$ and inversely proportional to its
mass $m_Q$.  Since $m_Q$ is large compared to the scale $\Lambda_{\rm
  QCD}$ of the light degrees of freedom (gluons and sea quarks) in the
hadron, hyperfine splittings in heavy-quark systems are small.  States
with approximately the same light-field content but differing only in
the relative spin states of the heavy quarks should therefore lie
close together in mass, an observation that lies at the crux of {\it
  heavy-quark spin symmetry}; any two such states related in this way
are said to differ only by a hyperfine splitting.

In the context of quarkonium with a heavy quark-antiquark pair $Q\bar
Q$ ($c\bar c$ or $b\bar b$), states related by hyperfine splittings
have the same wave functions with respect to the light degrees of
freedom, and differ in mass only due to the relative $Q$ or $\bar Q$
spin states.  In charmonium, for example, the $1S$ states ($\eta_c$,
$J/\psi$) form a hyperfine doublet, while the $1P$ states ($h_c, \,
\chi_{c0}, \, \chi_{c1}, \, \chi_{c2}$) form a hyperfine quartet.

The actual operators in a Hamiltonian formalism responsible for
hyperfine splittings are easy to identify by exploiting the
similarities of QED and QCD.  Essentially the same terms emerge from
the Breit reduction~\cite{Breit:1932} of Dirac fermions interacting
through photons and gluons~\cite{DeRujula:1975qlm}, respectively.
These terms are the well-known spin-spin, spin-orbit, and tensor
operators, as discussed in detail for quarkonium in
Ref.~\cite{Lebed:2017yme}.  The spin-spin operator $\bm{S}_Q \cdot
\bm{S}_{\bar Q}$ is especially interesting, because in the Breit
reduction it is accompanied by the Laplacian of the $1/r$ gauge
interaction, which produces the contact interaction
$\delta^{(3)}(\bm{r})$, a fact first noted by Fermi~\cite{Fermi:1930}.
Such a term contributes only for wave functions that are nonzero at
zero $Q {\bar Q}$ separation.  In the context of nonrelativistic
quantum mechanics, only the $S$ waves satisfy this criterion, and all
$L > 0$ states have a zero spin-spin hyperfine contribution.

A primary result of Ref.~\cite{Lebed:2017yme} is the observation that,
for any given hyperfine multiplet, one unique linear combination of
masses is sensitive solely to the spin-spin and not to the spin-orbit
or tensor operators.  In the case that all quarks are heavy, the
relevant combination for $L > 0$ states should therefore have a very
small mass splitting -- much smaller than typical hyperfine
splittings.  It is a linear combination of pairwise hyperfine
splittings and is therefore a hyperfine splitting in its own right,
dubbed {\it ultrafine\/} in Ref.~\cite{Lebed:2017yme}.  The ultrafine
combination is simply the difference between the mass of the state
with total quark spin $S_{Q\bar Q} = 0$ and the spin-averaged mass of
states with total quark spin $S_{Q\bar Q} = 1$.  This combination for,
{\it e.g.}, $P$-wave quarkonium is
\begin{equation} \label{eq:DeltaDef} \Delta \equiv M_h - \frac{1}{9}
  \left[ 1 \! \cdot \! M_{\chi_0} + 3 \! \cdot \! M_{\chi_1} + 5 \!
    \cdot \! M_{\chi_2} \right] \, .
\end{equation}

The ultrafine splittings were seen in all cases where all four states
have been observed to be extremely small---indeed, experimentally
consistent with zero---both in quarkonium~\cite{Lebed:2017yme} and
positronium~\cite{Lamm:2017lrn}.  The theoretically expected
splittings are so small that any measured deviation from zero can be
identified as the presence in at least some of the quarkonium states
of a substantial non-$Q {\bar Q}$ Fock state (``coupled-channel
exoticity'').  Such a component can be most easily identified with a
heavy-light meson pair [$(Q \bar q) (\bar Q q)$] contribution, but it
might also be due to a tetraquark or some other exotic component
reviewed in~\cite{Lebed:2016hpi}.

Essentially the same reasoning holds for the yet-unobserved
heavy-quark {\it hybrid mesons}, for which the gluon field connecting
the $Q \bar Q$ pair occurs in a nontrivial configuration (For a
review, see Ref.~\cite{Meyer:2015eta}).  Such states were discussed
briefly in Ref.~\cite{Lebed:2017yme}.  While of course no hybrid meson
has ever been experimentally confirmed as such, the existence of
excited gluon fields is a widely expected feature of QCD, and
moreover, a definite spectrum of heavy-quark hybrid mesons is a
well-established feature of increasingly sophisticated lattice QCD
simulations~\cite{Griffiths:1983ah,Juge:1997nc,Juge:2002br,
  Juge:1999ie,Bali:2000vr,Bali:2003jq,Liu:2012ze}.  In this paper we
investigate the hyperfine structure of heavy-quark hybrid mesons,
showing that all of the lowest-lying multiplets possess a combination
properly described as ultrafine, and that lattice simulations
definitively show these combinations to be generally smaller than
generic hybrid hyperfine splittings.

This paper is organized as follows: In Section~\ref{sec:BO} we briefly
review the Born-Oppenheimer formalism most convenient for describing
the hybrid states.  We review the heavy-quark description of hadrons
in Section \ref{sec:HQQCD}, and use it to argue that the hyperfine
interaction is short-ranged.  An analysis of the distinction between
hyperfine and ultrafine hybrid splittings is presented
Section~\ref{sec:Ultra}, and it is shown that matrix elements of the
hyperfine interaction are small in the heavy quark limit. The
hypothesis is then compared to recent lattice results in
Section~\ref{sec:Lattice}.  Conclusions appear in
Sec.~\ref{sec:Concl}.

\section{Hybrids in the Born-Oppenheimer Approximation}
\label{sec:BO}

Introduced nearly a century ago, the {\it Born-Op\-pen\-hei\-mer\/}
(BO) {\it approximation}~\cite{Born:1927boa} is based upon a scale
separation in atomic and molecular physics due to the heaviness of the
nuclei (mass $m_N$) compared to the electrons (mass $m_e$).  Even
though the same electromagnetic forces act upon both types of
constituent, the lighter electrons respond to these forces much more
quickly than do the heavier nuclei, and hence adiabatically adjust
their configuration to that of the comparatively static nuclei.
Quantitatively, the relative time scales associated with electronic
motion are shorter than those associated with nuclear motion by powers
of $m_e/m_N$.  Consequently, one may factor the full wave function of
the system into a part due entirely to the electronic configuration
and a part due entirely to the nuclei, which effectively act as static
electromagnetic sources.

In the BO approximation, the energy of the light degrees of freedom
interacting with the heavy degrees of freedom arrayed in a fixed
configuration defines a {\it Born-Oppenheimer potential}, which is
labeled both by the relative separations of the heavy sources and by
the symmetries exhibited by their configuration; in the case of a
homonuclear diatomic molecule, the BO potential depends only upon the
single nuclear separation $r$, and the potentials are labeled by the
irreducible representations of the group $D_{\infty h}$, which
describes the symmetries of a cylinder (coaxial with the nuclei).

These representations are labeled~\cite{Landau:1977} by the quantum
numbers $\Gamma \equiv \Lambda^\epsilon_\eta$, which are defined as
follows.  Starting with the total angular momentum $\bm{J}_{\rm
  light}$ of the light (electronic) degrees of freedom and the unit
vector $\hat {\bm{r}}$ connecting the nuclei, the eigenvalues $\lambda
= 0, \pm 1, \pm 2, \ldots$ of the axial angular momentum operator
$\hat {\bm{r}} \cdot \bm{J}_{\rm light}$ provide a good quantum
number; since the system is symmetric with respect to reflection
through any plane containing $\hat {\bm{r}}$ (which takes $\lambda \to
-\lambda$), energy eigenvalues cannot depend upon the sign of
$\lambda$, so that one defines $\Lambda \equiv |\lambda|$.
Analogously to the labels $S,P,D,\ldots$ for the usual angular
momentum quantum numbers $L=0,1,2,\ldots$, the values
$\Lambda=0,1,2,\ldots$ are then denoted by $\Sigma,\Pi,\Delta,\ldots$.
The light degrees of freedom also possess a reflection symmetry about
an origin given by the midpoint between the two nuclei, so that the
eigenvalue $\eta$ of the corresponding parity operator $P_{\rm light}$
is a good quantum number, with $+1 \, (-1)$ denoted by $g \, (u)$,
respectively.  Lastly, the $\Lambda = 0$ ($\Sigma$) representations
can be distinguished by their behavior under a reflection $R_{\rm
  light}$ through a plane containing the nuclei, its $\pm 1$
eigenvalue being denoted by $\epsilon$.  The $\Lambda > 0$
configurations $\left| \lambda, \eta; \bm{r} \right>$ can also be
combined into eigenstates of $R_{\rm light}$ with eigenvalue
$\epsilon$ by noting that $R_{\rm light} \left| \lambda, \eta; \bm{r}
\right> = (-1)^\lambda \zeta \left| -\lambda, \eta; \bm{r} \right>$,
where $\zeta$ is the intrinsic parity of the light degrees of freedom,
and by defining the eigenstates
\begin{equation}
  \left| \Lambda, \eta, \epsilon; \bm{r} \right> \equiv
  \frac{1}{\sqrt{2}} \left( \left| \Lambda, \eta; \bm{r} \right> +
    \epsilon (-1)^\Lambda \zeta \left| -\Lambda, \eta; \bm{r} \right>
  \right) \, .
\end{equation}

Full physical states for the system are then obtained by solving the
Schr\"{o}dinger equation of the nuclei interacting with the BO
potential $V_\Gamma(r)$, which produces eigenvalues and eigenfunctions
labeled by a principal quantum number $n$.  When the complete state
including both light and heavy components is considered, additional
good quantum numbers arise.  Components of $\bm{J}_{\rm light}$
orthogonal to $\hat{\bm{r}}$ are not among these, because the nuclei
can possess their own relative orbital angular momentum $\bm{L}_{\rm
  nuc}$ (which satisfies $\hat {\bm{r}} \cdot \bm{L}_{\rm nuc} = 0$)
that cannot be distinguished from $\bm{J}_{\rm light}$ in the complete
state; the conserved quantity is the combined orbital angular momentum
$L$, where
\begin{equation} \label{eq:Ldef}
\bm{L} \equiv \bm{L}_{\rm nuc} + \bm{J}_{\rm light} \, ,
\end{equation}
as is the total nuclear spin $S$, and ultimately, the total angular
momentum quantum numbers $J, J_z$ of the molecule.  It is also easily
seen from contracting Eq.~(\ref{eq:Ldef}) with $\hat {\bm{r}}$ that $L
\ge |\hat {\bm{r}} \cdot \bm{L}| = |\hat {\bm{r}} \cdot
\bm{J}_{\rm light}| = \Lambda$.  In summary, the physical states are
completely specified by the kets
\begin{equation}
\left| n, L, S, J m_J; \Lambda, \eta, \epsilon \right> \, ,
\end{equation}
with $J_{\rm light}$ and $L_{\rm nuc}$ eigenvalues implicit.

The same BO approach can be applied to any other quantum-mechanical
system that possesses both heavy and light degrees of freedom.  In the
context of heavy quarkonium $Q \bar Q$, all the nomenclature discussed
thus far in this section applies verbatim with small
substitutions~\cite{Braaten:2014qka}: The nuclei become the $Q \bar Q$
pair, $\hat {\bm{r}}$ is the unit vector pointing from $\bar Q$ to
$Q$, the light electronic degrees of freedom become the glue field,
and the presence of the antiparticle $\bar Q$ means that $\eta$ is the
eigenvalue not of $P_{\rm light}$ [readily seen to equal $\epsilon
(-1)^\Lambda$ in this case], but $(CP)_{\rm light}$.  A {\it hybrid
  heavy quarkonium meson\/} in this terminology then simply refers to
an energy eigenstate containing $Q\bar Q$ for which at least some of
the eigenvalues $\Lambda, \eta, \epsilon$ are nontrivial, {\it i.e.},
every BO eigenstate except $\Sigma^+_g$.  Taking into account both the
light and heavy degrees of freedom, the overall discrete quantum
numbers for the physical state are determined to be:
\begin{eqnarray}
P & = & \; \; \epsilon (-1)^{\Lambda + L + 1} \, , \label{eq:Pval} \\
C & = & \eta \epsilon (-1)^{\Lambda + L + S} \, . \label{eq:Cval}
\end{eqnarray}

The first treatment of hybrids within the BO approximation (and
moreover, in the context of a lattice QCD simulation) appeared several
decades ago in Ref.~\cite{Griffiths:1983ah}.  In
Sec.~\ref{sec:Lattice} we touch upon important landmarks in the
lattice simulations of hybrids and their connection to the BO
approximation.  Here however, we especially note two important recent
papers in this regard: First, Ref.~\cite{Braaten:2014qka} established
the BO approximation as a formalism useful not only for the
description of hybrid mesons, but the full collection of exotic
$XY\!Z$ states~\cite{Lebed:2017yme} as well; as noted
in~\cite{Braaten:2014qka}, the light degrees of freedom can be
generalized to carry nontrivial isospin quantum numbers and therefore
can also be used to study tetraquarks.

Moreover, Ref.~\cite{Braaten:2014qka} noted (an observation dating
back as far as~\cite{deg}) that some of the hybrid BO
light-field potentials become degenerate in the $r \to 0$ limit, in
which case the light configuration (specifically, one in the
color-adjoint representation) is called a {\it gluelump}.  Given
eigenvalues of $L$ and $\Lambda$ satisfying $L \ge 1$, $L > \Lambda$,
as well as eigenvalues of $\eta$, $\epsilon$, and $n$, and using
Eqs.~(\ref{eq:Pval})--(\ref{eq:Cval}), one finds that the quartet of
states (one from $S=0$ and three from $S=1$) derived from the
$\Lambda^\epsilon_\eta (nL)$ and $(\Lambda + 1)^{-\epsilon}_\eta (nL)$
BO potentials produce the same set of $J^{PC}$ values.  The lowest BO
potentials above the ground state $\Sigma^+_g$ are calculated on the
lattice to be $\Sigma^-_u (1P)$, $\Pi^+_u (1P)$, and $\Pi^-_u (1P)$
(see Sec.~\ref{sec:Lattice}).  The potentials $\Pi^{\pm}_u (1P)$ form
a parity pair expected to be degenerate in the large-$m_Q$ ({\it
  i.e.}, leading-order BO) limit.  However, the pair $\Sigma^-_u (1P)$
and $\Pi^+_u (1P)$, an example of the above rule, each produce a set
of states with the same $J^{PC}$ values (see Table~\ref{tab:Lattice}),
and indeed arise from the same gluelump, $J_{\rm light}^{P_{\rm light}
  C_{\rm light}} = 1^{+-}$.

The second important paper~\cite{Berwein:2015vca} built upon the BO
heavy-quark hybrid studies of Ref.~\cite{Braaten:2014qka} to develop
an effective theory in the expansion parameter $1/m_Q$, in order to
study such states.  As shown in~\cite{Berwein:2015vca}, the
$\Sigma^-_u (1P)$ and $\Pi^+_u (1P)$ potentials explicitly produce
coupled Schr\"{o}dinger equations that, in particular, lift the
$\Pi^{\pm}_u (1P)$ degeneracy, an effect known in BO studies as
$\Lambda$-{\it doubling}.  Reference~\cite{Berwein:2015vca} then
showed how to diagonalize and numerically solve the Schr\"{o}dinger
equations, thus obtaining a spectrum of heavy-quark hybrid meson
masses that compare favorably with the results of lattice simulations.

The precise choice for an expansion parameter in the heavy-quarkonium
hybrid sector to obtain a useful BO expansion corresponds to the
nature of the optimal effective field theory used to describe the
states.  In a very recent paper~\cite{Brambilla:2017uyf} by several of
the authors of Ref.~\cite{Berwein:2015vca}, it is argued that the
hierarchy  relevant to hybrids is
\begin{equation} \label{eq:EFT}
m_Q v \gg \Lambda_{\rm QCD} \gg m_Q v^2 \, ,
\end{equation}
where $v$ is the typical heavy-quark velocity ($\sim \alpha_s \approx
0.3$ for charm).  This choice is motivated by requiring that the
typical Bohr radius-like heavy-quark separation, $1/(m_Q v)$, is small
compared to the size $1/\Lambda_{\rm QCD}$ of the light hadronic
cloud, and that the typical heavy-quark kinetic energies are small
compared to those of the light cloud.  With $m_c \approx 1.5$~GeV and
$\Lambda_{\rm QCD} \approx 300$~MeV, Eq.~(\ref{eq:EFT}) becomes
450~MeV $\gg$ 300~MeV $\gg$ 130~MeV\@.  A hadronic BO expansion
would nevertheless still have been possible even under the weaker
condition that $\Lambda_{\rm QCD}/m_Q$ ($\approx 0.2$ for charm) is a
small parameter.

\section{Heavy-Quark QCD}
\label{sec:HQQCD}

The splittings that occur in heavy-quark hybrid multiplets are
determined by the spin-dependent structure of QCD, which can be 
determined by constructing the heavy-quark expansion of the
Hamiltonian of QCD in Coulomb gauge (a convenient choice because all
degrees of freedom are physical)~\cite{Szczepaniak:1996tk}.  A
gauge-invariant approach based upon the Wilson loop yields the same
results~\cite{Eichten:1980mw}.

The first spin-dependent interaction arises at order $1/m_Q$ and is
generated by the Foldy-Wouthuysen term

\begin{eqnarray}
\mathcal{H}_1 &=& \frac{1}{2m_Q} \int d^3 x \, h^\dagger({\bf x})
\left({\bf D}^2 - g \bm{\sigma} \cdot
{\bf B} \right) h({\bf x}) \nonumber \\
&& - ( h \rightarrow \chi; \, m_Q \rightarrow m_{\bar Q}) .
\label{eq:1m}
\end{eqnarray}
Here, $h$($\chi$) annihilates a heavy (anti)quark, ${\bf B}$ is the
chromomagnetic field, and ${\bf D}$ is the covariant
derivative. Spin-dependence is carried by the matrix element of
$\bm{\sigma}\cdot {\bf B}$ and is zero in conventional mesons because
the only available vector is ${\bf r} = \bf{r}_Q - \bf{r}_{\bar Q}$,
which has the wrong parity to yield a nonzero
result~\cite{Eichten:1980mw}. However, this conclusion need not follow
in the case of hybrids where additional vectors, such as $\bm{J}_{\rm
light}$, can contribute (see Ref.~\cite{Oncala:2017hop}, Eq.~(18) and
subsequent discussion).  The effect this interaction has on the
ultrafine splitting will be discussed below.

The sole spin-dependent term at order $1/m_Q^2$ is given by
\begin{eqnarray}
\mathcal{H}_2 &=& \frac{1}{8 m_Q^2} \int d^3 x\, h^\dagger({\bf x})
 g \bm{\sigma}\cdot ({\bf E} \times {\bf D} - {\bf D} \times {\bf E})
\, h({\bf x})  \nonumber \\
&& + (h \rightarrow \chi; \, m_Q \rightarrow m_{\bar Q}).
\end{eqnarray}
The chromoelectric field is denoted ${\bf E}$ in this expression.
Standard perturbation theory and some manipulation yields the
classical and Thomas precession portions of the spin-orbit
interaction~\cite{Szczepaniak:1996tk}:
\begin{eqnarray}
\lefteqn{V^{\rm cLS}_{\Gamma}({\bf r}=
  {\bf r}_Q-{\bf r}_{\bar Q})} && \nonumber \\
& = & 
  \left( \frac{\bm{\sigma}_Q \cdot {\bf L}_Q }{ 4 m_Q^2} -
    \frac{\bm{\sigma}_{\bar Q} \cdot {\bf L}_{\bar Q} }
    { 4 m_{\bar Q}^2} \right)
  \frac{1}{r}
  \frac{d V_{\Gamma} }{d r} \, ,
\end{eqnarray}
where $V$ is the static (Wilson-loop) interquark potential.
Dependence upon the gluonic adiabatic quantum numbers $\Gamma =
\Lambda^\epsilon_\eta$ has been made explicit here, and reveals that
the classical spin-orbit interaction is fixed by the relevant
adiabatic potential.  Since hybrid potentials are relatively flat at
distance scales around 1~fm (corresponding to their expected
equilibrium size), one concludes that the classical spin-orbit
contribution to hybrid mass splittings is smaller than for
conventional mesons.

Additional spin splittings arise at order $1/m_Q^2$ by iterating the
first-order terms in the Foldy-Wouthuysen expansion
[Eq.~(\ref{eq:1m})].  In particular, the expression involving two
powers of ${\bm \sigma} \cdot {\bf B}$ gives rise to the (generalized)
hyperfine interaction:
\begin{eqnarray}
  \lefteqn{V^{\rm hyp}_{\Gamma}
    ({\bf r}={\bf r}_Q-{\bf r}_{\bar Q})
    = \alpha_s \frac{4\pi}{3 m_Q\, m_{\bar Q}}\, {\bf S}_f
    \! \cdot {\bf S}_{\bar f}} && \nonumber \\
  & \times & \sum_{\Gamma^\prime 
    \neq \Gamma}
  \frac{1}{E_{\Gamma}(r) -
    E_{\Gamma^\prime}(r)} \nonumber \\
  & \times & \left< \Gamma ; {\bf r}_Q ,
    {\bf r}_{\bar Q} \left|
      \int d^3x \, h^\dagger({\bf x}) {\bf B}({\bf x}) h({\bf x})
    \right| \Gamma^\prime; {\bf r}_Q , {\bf r}_{\bar Q}
  \right> \nonumber \\
  & \cdot & \left< \Gamma^\prime; {\bf r}_Q ,
    {\bf r}_{\bar Q} \left|
      \int d^3y \, \chi^\dagger({\bf y}) {\bf B}({\bf y})
      \chi({\bf y})
    \right| \Gamma ; {\bf r}_Q , {\bf r}_{\bar Q}
  \right>
  \nonumber \\
  & + & (h \leftrightarrow \chi) \, .
\label{eq:V4}
\end{eqnarray}
The energy denominator is expressed in terms of adiabatic energies as
a function of the distance $r$ between the color source and sink and
of the quantum numbers $\Gamma$ of the relevant adiabatic surface.
Notice that the intermediate state is summed over all adiabatic
surfaces that are coupled to the initial surface by a chromomagnetic
field.

We now argue that the spin-spin hyperfine interaction should be
short-ranged ($r \ll 1/\Lambda_{\rm QCD}$).\footnote{The expectation
of a short-ranged hyperfine interaction for the lowest bound-state
configurations is supported by observed ultrafine splittings in heavy
quarkonia and in positronium~\cite{Lebed:2017yme}.}  First, recall
that in Coulomb gauge the vector potential propagates over short
distances, in distinction to the instantaneous interaction that gives
rise to confinement.  Perturbatively, the matrix element leading order
is proportional to two derivatives acting upon $1/r$, which yields a
$\delta^{(3)}(\bm{r})$; its leading perturbative corrections are
presented in~\cite{Penin:2004xi}.  Nonperturbatively, one expects this
behavior to be replaced by two derivatives acting on a Yukawa-like
interaction (its mass scale appearing because the primary
nonperturbative effect in QCD is the generation of a mass gap),
although the long-range string potential produces an additional term
proportional to $1/r^5$~\cite{Brambilla:2014eaa}.  Each also yields a
short-range matrix element.

This expectation is supported by lattice measurements of the
ground-state hyperfine interaction in quenched-lattice QCD, which
indicate that the potential is zero within statistical errors for $r >
0.2$~fm~\cite{Koma:2006fw}.  This result implies that the hybrid
hyperfine interaction is also short-ranged, by the following argument.
The lattice computation of the hybrid static potential is made by
placing the gluonic source and sink into nontrivial configurations.
These gluonic configurations differ from the ground-state ones only up
to distance scales of some fraction of a fm; thus, one expects the
ground-state and excited-state static potentials to differ by no more
than a constant at large distances.  In fact, the slopes of all
measured static potentials are the same past approximately
2~fm~\cite{Juge:1997nc}, while their differences at large distances
are known from the strict QCD string picture to scale as
$1/r$~\cite{Luscher:2004ib}.  Measurements of the spin-spin
interaction are obtained by inserting operators on the temporal legs
of the relevant Wilson loops~\cite{Koma:2006fw}.  Because these are
short-distance operators, and because their matrix elements are
observed to decorrelate in the ground-state Wilson loop at large
distance, they are also expected to decorrelate in the excited-state
Wilson loop at large distance.  We shall see shortly that this
expectation is confirmed in heavy-quark lattice data.

\section{Ultrafine Hybrid Splittings}
\label{sec:Ultra}

The general definition of the ultrafine mass splitting for a set of
states in the same multiplet (hence with the same BO potential
$\Lambda^\epsilon_\eta$, principal quantum number $n$, and orbital
quantum number $L > 0$) is simply the spin-averaged difference between
the states with total heavy-quark spin $S=0$ and $S=1$, and is easily
seen to be~\cite{Lebed:2017yme}:
\begin{eqnarray}
\Delta_{n,L} & \equiv & M(n{}^1 \! L_{J=L}) \nonumber \\
& - & \frac{2L-1}{3(2L+1)} M(n{}^3 \! L_{J = L-1}) \nonumber \\
& - & \frac{2L+1}{3(2L+1)} M(n{}^3 \! L_{J = L}) \nonumber \\
& - & \frac{2L+3}{3(2L+1)} M(n{}^3 \! L_{J = L+1}) \, ,
\label{eq:HyperGen}
\end{eqnarray}
of which Eq.~(\ref{eq:DeltaDef}) is merely the $L=1$ case for ordinary
quarkonium.  The expression for $L=0$ ($S$ waves, hence consisting
solely of a $\Sigma$ BO potential) is even simpler: $\Delta_{n,0}
\equiv M(n{}^1 \! S_0) - M(n{}^3 \! S_1)$.

We now turn to the issue of whether the ultrafine combination is truly
unique in a given hybrid hyperfine multiplet.  After all, hybrid
quarkonia are more complicated states than conventional quarkonia; in
the latter case, the only available operators to form Hamiltonian
terms that can split the hyperfine multiplet are the heavy-quark spins
$\bm{S}_Q$, $\bm{S}_{\bar Q}$, which are coupled to each other and to
the sole orbital angular momentum operator $\bm{L}$ and the $Q\bar Q$
direction $\hat{\bm{r}}$ to form the well-known
operators~\cite{DeRujula:1975qlm}:
\begin{eqnarray}
\label{eq:hyperfine}
\bm{S}_Q \! \cdot \! \bm{S}_{\bar Q} && {\rm (hyperfine)} \, , \\
\label{eq:spinorbit}
{\bm S} \! \cdot \! \bm{L} &&
{\rm (\mbox{spin-orbit})} \, , \\
\label{eq:tensor}
\overset{\text{\tiny$\bm\leftrightarrow$}}{T}
\equiv (\bm{S}_Q \! \cdot \hat{\bm{r}})
(\bm{S}_{\bar Q} \! \cdot \hat{\bm{r}})
-\frac 1 3 \bm{S}_Q \! \cdot \! \bm{S}_{\bar Q} && {\rm (tensor)} \, ,
\end{eqnarray}
where ${\bm S} \equiv \bm{S}_Q \! + \! \bm{S}_{\bar Q}$.  As noted
in~\cite{Lebed:2017yme}, the contribution of the latter two operators
to the ultrafine combination Eq.~(\ref{eq:HyperGen}) vanishes for
group-theoretical reasons: They have rank $>$ 0 in $\bm{L}$ space,
while Eq.~(\ref{eq:HyperGen}) includes multiplets in $\bm{L}$ space
only in the rank-0 (symmetric) combination.

The same argument applies equally for the corresponding operators in
the case of hybrid states: The matrix elements of
Eqs.~(\ref{eq:spinorbit}) and (\ref{eq:tensor}) vanish when the full
orbital angular momentum ${\bm L}$ is used, leaving only the spin-spin
operator of Eq.~(\ref{eq:hyperfine}).  However, now one must consider
additional possible operators.  One begins by noting that the analogue
of $\bm{L}$ in the conventional case (the operator dictating the
short-distance behavior of the wave function) is $\bm{L}_{Q\bar Q}$ in
the hybrid case which, like $\bm{J}_{\rm light}$, does not provide
good quantum numbers.  Nevertheless, we have noted above that all the
lightest hybrid multiplets $H_{1,2,3,4}$ arise from the same $J_{\rm
light} = 1$ gluelump, while $H_1$ has $L_{Q\bar Q}=0$ and $H_{2,3,4}$
have $L_{Q\bar Q} = 1$.  Inasmuch as each $H_i$ corresponds to a
unique BO potential with a good $L$ quantum number, states of the same
$J^{PC}$ in different $H_i$ (specifically, $1^{+-}$ and $2^{+-}$) do
not mix; however, the presence of an operator like $\bm{S} \cdot
\bm{L}_{Q\bar Q}$ could accomplish this mixing.  Within a given BO
potential, however, one can check explicitly (by expanding all states
in terms of eigenstates of the operator $\bm{J}_{Q\bar Q}
\equiv \bm{S} + \bm{L}_{Q\bar Q}$) that contributions to
Eq.~(\ref{eq:HyperGen}) by spin-orbit or tensor operators containing
$\bm{L}_{Q\bar Q}$ or $\bm{J}_{\rm light}$ cancel.

As discussed in Sec.~\ref{sec:HQQCD}, a spin-dependent interaction can
occur in hybrids at order $1/m_Q$ due to the availability of
additional spin vectors.  Such contributions appear in the mixing of
hybrids with quarkonium, as well as among hybrid states (see Eq.~(14)
in Ref.~\cite{Soto:2017one}).  Even so, the ultrafine splittings stand
out as the special case of hyperfine splittings that receive a nonzero
contribution {\em only\/} from the heavy-quark spin-spin coupling;
indeed, using the results from analyzing Eq.~(14)--(17) of
Ref.~\cite{Soto:2017one}, one can show explicitly that our ultrafine
combination Eq.~(\ref{eq:HyperGen}) vanishes.

On the other hand, the mixing of the BO potentials $\Sigma^-_u (1P)$
and $\Pi^+_u (1P)$ in $H_1$ leaves $L$ (=1) invariant, and indeed, the
mixing between all states in these configurations derived
in~\cite{Berwein:2015vca} depends solely upon $L$ (and similarly for
$L=2$ in $H_4$).  Therefore, the mixing in this case does not spoil
the cancellation in Eq.~(\ref{eq:HyperGen}) of matrix elements of
Eqs.~(\ref{eq:spinorbit})--(\ref{eq:tensor}) for the states in $H_1$.

The question of whether the combination Eq.~(\ref{eq:HyperGen}) is
ultrafine for a given $H_i$ thus comes down to whether the spatial
wave function associated with the BO potential vanishes as $r \to 0$.
One of course anticipates the usual $r^L$ behavior so that only
$S$-wave multiplets fail to have an ultrafine splitting, but the BO
potentials exhibit some interesting quirks.

First, the multiplets $H_2$ and $H_4$ are $P$-wave and $D$-wave,
respectively, so that one expects each multiplet to have an ultrafine
splitting $\Delta$ defined by Eq.~(\ref{eq:HyperGen}).  Indeed, using
the lattice values in Table~\ref{tab:Lattice}, one finds $\Delta$ to
be consistent with zero for $H_2$ and to differ from zero by only
1.3$\sigma$ for $H_4$.  Note also that the largest hyperfine
splitting, $D$, within these multiplets is only about a factor two
larger in $H_2$ and $H_4$.  The pattern of splittings in these
multiplets is quite peculiar; in all the known quarkonium and
positronium cases~\cite{Lebed:2017yme}, the spin-triplet states have
masses that increase monotonically with $J$, and the spin-singlet
state lies between the lowest and highest triplet state.  However, in
$H_2$ the $1^{+-}$ triplet state lies below the $0^{+-}$ triplet
state, and the $1^{++}$ singlet state lies above all the triplet
states.  In $H_4$, the $2^{++}$ singlet state lies below all the
triplet states.  One may attribute these peculiarities to the size of
lattice uncertainties, neglect of mixing between $1^{+-}$ or $2^{+-}$
states between BO multiplets, or misidentification of lattice states
with the correct BO multiplets.  The smallness of the ultrafine
splitting in both cases, however, argues that none of these
possibilities need be true, and that the calculated hybrid spectrum
ordering is indeed correct; refined lattice simulations of the
splittings would certainly serve to clarify the situation further.

Second, $H_3$ is an $S$-wave multiplet, suggesting a nonzero wave
function as $r \to 0$, but its $(CP)_{\rm light}$ and $R_{\rm light}$
quantum numbers are both $-1$; and being a $\Lambda = 0$ ($\Sigma$)
state, it has $P_{\rm light} = -1$ as well, suggesting a wave function
that vanishes at the origin.  The only way to reconcile these facts is
to allow the wave function to be odd, changing sign discontinuously
when passing through the origin.  In that case, integrating over the
symmetric $\delta^{(3)}(\hat{\bm{r}})$ distribution gives a vanishing
result.  Indeed, the splitting Eq.~(\ref{eq:HyperGen}) for $H_3$ from
Table~\ref{tab:Lattice} is extremely small, suggesting an ultrafine
splitting.

Lastly, let us turn to the lowest multiplet, $H_1$.  In this case, the
ordering of the masses in Table~\ref{tab:Lattice} seems completely
conventional, and both BO potentials are $P$-wave, suggesting a
noncontroversial ultrafine splitting with a wave function $\sim r^1$
as $r \to 0$.  Indeed, the value for $\Delta$ (especially compared to
the largest intermultiplet splitting $D$) appears to confirm this
suspicion.  However, in this case the mixing of BO potentials
discussed in~\cite{Berwein:2015vca} generates normalizable wave
functions with asymptotic behavior $r^{L-1}$ and $r^{L+1}$.  In
particular, since $L=1$ in this case, one finds a wave function
component that survives as $r \to 0$!  However, the angular part of
the wave function is still one that corresponds to $L=1$.  The general
expression for these angular wave functions is given
in~\cite{Berwein:2015vca}, and indeed, in textbooks as
well~\cite{Landau:1977}; in the case $\Lambda = 0$ they reduce to the
usual spherical harmonics.  The important point, however, is that only
the $L = \Lambda = 0$ angular wave function has trivial angular
dependence and hence is well defined at the origin, meaning that
again, the full wave function changes sign at the origin and offers
zero support to the symmetric $\delta^{(3)} (\hat{\bm{r}})$
distribution.  The splitting of Eq.~(\ref{eq:HyperGen}) for $H_1$ is
therefore indeed ultrafine.

\section{Hybrids in Lattice QCD Simulations}
\label{sec:Lattice}

In the absence of confirmed experimental evidence for hybrid mesons,
one may rely upon the direct results of numerical simulations of QCD
on a discretized lattice.  As noted above, the first lattice
simulation to make use of the BO approximation for hybrids appeared
almost 35 years ago~\cite{Griffiths:1983ah}.  The first high-quality
determinations of the BO potentials relevant to the heavy-quark
hybrids were performed in Ref.~\cite{Juge:1997nc}, with computations
on larger lattice volumes presented in Ref.~\cite{Juge:2002br}.  Some
details of the hybrid spectrum were also discussed in
Ref.~\cite{Juge:1997nc}, with further improvements in
Ref.~\cite{Juge:1999ie}.  These calculations were carried out in the
quenched limit, the first unquenched simulations~\cite{Bali:2000vr}
(with an equivalent pion mass of 650~MeV) giving very similar results.

Already noted in the early work~\cite{Juge:1997nc} was the
near-degeneracy of $\Sigma^-_u$ and $\Pi_u$ potentials in the $r \to
0$ limit.  Simulations in Ref.~\cite{Bali:2003jq} used a finer lattice
spacing to explore this short-distance regime.  The potentials appear
to approach a single $J_{\rm light}^{P_{\rm light} C_{\rm light}} =
1^{+-}$ gluelump energy, which was first calculated
in~\cite{Foster:1998wu}.

Improved calculations of the gluelump spectrum appeared rather
recently in Ref.~\cite{Marsh:2013xsa}.  Notably, all lattice
simulations agree that the lightest gluelump has quantum numbers
$1^{+-}$, with the first and second excited gluelumps being $1^{--}$
and $2^{--}$ states, respectively.  Interestingly, $1^{+-}$ and
$1^{--}$ are the quantum numbers of a chromomagnetic and
chromoelectric constituent gluon, respectively; however, the lattice
approach is intrinsically nonperturbative, meaning that one should
have no expectation for a constituent approach to apply here.

Currently, the best lattice simulations of the heavy-quark hybrid
spectrum (specifically, charmonium) have been produced by the Hadron
Spectrum Collaboration~\cite{Liu:2012ze}.  The calculation was
unquenched, with an effective pion mass of 400~MeV, although it should
be noted that explicit meson-meson operators were not included in the
simulation.  In practice, this omission means that long-distance
light-quark effects are not present in the lattice hybrid spectrum,
and thus small ultrafine splittings are expected.  The effort produced
46 states in the charmonium sector, spread over 17 distinct $J^{PC}$
channels.  Of these, of course all the lowest conventional $c\bar c$
states appear, but so do a number of states that fill complete hybrid
multiplets.  The two types of states are distinguished on the lattice
by identifying the dominant interpolating operator in their
construction: If the covariant QCD derivatives contribute primarily
through the ordinary derivative part, the states are identified as
conventional $c\bar c$; if they contribute primarily through their
commutators, {\it i.e.}, the QCD field strength, the states are
identified as hybrids.

In the context of Ref.~\cite{Liu:2012ze}, the hybrids are organized
according to supermultiplets one would obtain from the constituent
gluon model, starting with a $1^{+-}$ gluelump: a lower multiplet (4
states) based upon $L_{Q\bar Q} = 0$, and a higher multiplet (10
states) based upon $L_{Q\bar Q} = 1$.  The requisite states were all
observed in the simulation, and their masses (with respect to that of
the $\eta_c$) were determined; see Table~\ref{tab:Lattice}.
Significantly, all of the states emerge from the {\em same\/} $1^{+-}$
gluelump.

Upon examining the results of Ref.~\cite{Liu:2012ze} in detail,
Ref.~\cite{Braaten:2014qka} noted that the supermultiplets actually
divide into complete BO multiplets $H_{1,2,3,4}$, as labeled in
Table~\ref{tab:Lattice} (where we include the mixing in $H_1$ and in
$H_4$ advocated by Ref.~\cite{Berwein:2015vca}).  The analysis
of~\cite{Braaten:2014qka} using the numerical results
of~\cite{Liu:2012ze} showed that one can identify a genuine
organization of the states by mass values into these multiplets,
although the mass splitting between the highest state of one multiplet
and the lowest state of the next can sometimes be small, as one can
see in Table~\ref{tab:Lattice}.  It is important to note that the
method for distinguishing $H_{1,2,3,4}$ multiplets uses a somewhat
less general theoretical approach than that for distinguishing between
$c\bar c$ and hybrid states or between states of different
$J^{PC}$---it depends upon identifying specific chromoelectric and
chromomagnetic operators in the multipole expansion of the gluon field
with specific BO potentials~\cite{Brambilla:1999xf}---but the expected
patterns definitely hold.  Note also that the spectrum of $L_{Q\bar Q}
= 0$ states matches that of $H_1$, while the spectrum of $L_{Q\bar Q}
= 1$ states matches that of $H_2 \cup H_3 \cup H_4$, that the two sets
have no states of the same $J^{PC}$ in common, and that $H_{2,3,4}$
all have distinct $L$ values.

Indeed, the statistical uncertainties are sufficiently small that one
may use them to explore the spin structure of hybrid multiplets, as is
done here.  One must take caution to note that the uncertainties do
not take into account the extrapolation to physically small
light-quark masses, nor to the continuum limit of the lattice.
However, one may argue that the {\em differences\/} of masses should
be less sensitive to these effects than their absolute values (indeed,
one may suspect the same argument to hold for some portion of the
statistical uncertainties).  In any case, we assume that lattice
simulations of the hybrids are now sufficiently mature that one can at
last make definitive statements about their mass splittings.

\begin{table}[h]
\begin{tabular*}{\columnwidth}{l@{\extracolsep{\fill}}cccc} \hline \hline
\;\;\;\;\;\; \text{Multiplet}& $J^{PC}$ & $m \,$(MeV) & $D \, $(MeV)
& $\Delta \,$(MeV) \\ \hline
$H_1$ [$\Sigma^-_u (1P), \Pi^+_u (1P)$]
       & $1^{--}$ & $4285(14)$ & $139(21)$ & $5.4(17.8)$ \\
       & $0^{-+}$ & $4195(13)$ & & \\
       & $1^{-+}$ & $4217(16)$ & & \\
       & $2^{-+}$ & $4334(17)$ & & \\ \hline

$H_2$ [$\Pi^-_u (1P)$]
       & $1^{++}$ & $4399(14)$ &  $55(40)$ & $22(29)$ \\
       & $0^{+-}$ & $4386(09)$ & & \\
       & $1^{+-}$ & $4344(38)$ & & \\
       & $2^{+-}$ & $4395(40)$ & & \\ \hline

$H_3$  [$\Sigma^-_u (1S)$]
       & $0^{++}$ & $4472(30)$ &   $5(36)$ & $-5(36)$ \\
       & $1^{+-}$ & $4477(19)$ & & \\ \hline

$H_4$  [$\Sigma^-_u (1D), \Pi^+_u (1D)$]
       & $2^{++}$ & $4492(21)$ &  $56(30)$ & $-33(25)$ \\
       & $1^{+-}$ & $4497(39)$ & & \\
       & $2^{+-}$ & $4509(18)$ & & \\
       & $3^{+-}$ & $4548(22)$ & & \\ 
\hline\hline
\end{tabular*}
\caption{Charmonium hybrid masses from lattice QCD simulations by the
  Hadron Spectrum Collaboration~\cite{Liu:2012ze}, as adapted from
  Ref.~\cite{Berwein:2015vca} (where the experimental value of
  $m_{\eta_c}$ is added).  Also presented are the maximum mass
  difference $D$ and the ultrafine combination $\Delta$ within each
  multiplet.}
\label{tab:Lattice}
\end{table}

These arguments lead us to expect, on general and essentially
model-independent grounds, that the ultrafine splitting in hybrid
heavy-quark multiplets should be dominated by the matrix element of
$V^{\rm hyp}_{\Gamma}$, which in turn should be very small in the case
of hybrids, provided there is little wave-function support at the
origin.  We expect splittings that are much smaller than $\Lambda_{\rm
  QCD}$, and in practice on the order of an MeV or less.  This
conclusion is predicated on the applicability of the heavy-quark
expansion and the assumption that valence light-quark degrees of
freedom are negligible for hybrids.  A large measured ultrafine
splitting would thus be a very strong indication of the presence of
``coupled-channel exoticity'' in the multiplet, just as for
conventional quarkonium~\cite{Lebed:2017yme}.

\section{Conclusions}
\label{sec:Concl}

The QCD Hamiltonian specifies a limited number of spin-dependent
interactions that can serve to split heavy-quark multiplets of mesons,
either conventional or with excited gluonic degrees of freedom.  In
the absence of substantial light valence-quark degrees of freedom, the
ultrafine splitting defined in Eq.~(\ref{eq:HyperGen}) is expected to
be small on quite general grounds.  This observation holds in the case
of positronium, $1P$ and $2P$ bottomonium, and $1P$
charmonium~\cite{Lebed:2017yme}.  We anticipate that the ultrafine
splitting will \textit{not} be small for $2P$ charmonium because of the widely
accepted notion that the $X(3872)$ is not a pure $c\bar c$ state.

The expected small hyperfine splittings in hybrid multiplets will
assist in interpreting future spectroscopic data concerning hybrid
mesons, such as at PANDA.  As with conventional mesons, significant
ultrafine splittings will constitute essentially model-independent
evidence for coupled-channel exoticity in the relevant hybrid
multiplet.  The magnitude of an ultrafine splitting that qualifies as
``significant'' can be estimated from the charmonium $2P$ multiplet,
assuming that the $X(3872)$ is purely an interloper state.  Indeed,
the $1P$ charmonium ultrafine splitting is measured to be
$\Delta_{1,P} = 80 \pm 130$~keV, while typical quark models ({\it
  e.g.}, Ref.~\cite{Barnes:2005pb}) indicate that the $X$ is
approximately 100 MeV lighter than expected, hence $\Delta_{2,P} =
20$--$30$~MeV\@.  Thus, as a rough guide, ultrafine splittings that
are larger than 1--10~MeV are indicative of non-conventional valence
content in at least one state of a heavy-quark multiplet.

A likely minimal condition for the presence of coupled-channel
exoticity is the existence of nearby $S$-wave continuum thresholds.
We therefore examine the possibility of large ultrafine splittings in
the hybrids of Table~\ref{tab:Lattice} by mixing with open-charm $S$-
and $P$-wave meson pairs.  Positive-parity states can be made from
$SS$ or $PP$ combinations.  The former start at 3740~MeV for $D\bar
D$~\cite{Olive:2016xmw}, and run to 4220~MeV for $D_s^* \bar D_s^*$,
all of which are lighter than the ``bare'' positive-parity hybrids of
$H_2 \cup H_3 \cup H_4$ in Table~\ref{tab:Lattice}.  Alternatively,
the $PP$ combinations start at 4640~MeV ($D^*_{s0} \bar D^*_{s0}$),
and are therefore too heavy to create substantial light-quark valence
degrees of freedom in the positive-parity hybrids.  Thus we (rather
naively) expect the multiplets $H_2$, $H_3$, and $H_4$ to have very
small ultrafine splittings.  Negative-parity channels can be
constructed from $SP$ meson combinations.  Of these, $D\bar D_2^*$
lies close to the $J^{PC}=2^{-+}$ $H_1$ state, while $D\bar D_1$ lies
close to the $1^{--}$ $H_1$ state.  Thus, one sees an intriguing
possibility of a large ultrafine splitting in the lightest ($H_1$)
multiplet.

\begin{acknowledgments}
  \vspace{-2ex} This work was supported by the National Science
  Foundation under Grant No.\ PHY-1403891.
\end{acknowledgments}


\end{document}